\documentclass{mn2e}
\usepackage{psfig}
\usepackage{mnras_cite}
\newcommand{\redshiftnine}{\left( \frac{1+z}{10} \right)}

\newcommand{\geqsim}{\,\raisebox{-0.6ex}{$\buildrel > \over \sim$}\,}

\begin{document}

\title[Foregrounds for 21cm Observations of Neutral Gas at High
  Redshift]{Foregrounds for 21cm Observations of Neutral Gas at High
  Redshift}   
\author[Oh \& Mack]{S. Peng Oh \& Katherine J. Mack\\
Theoretical Astrophysics, Mail Code 130-33, Caltech, Pasadena, CA
  91125, USA}

\maketitle

\begin{abstract}
We investigate a number of potential foregrounds for an ambitious goal of future radio telescopes such as the Square
Kilometer Array (SKA) and Low Frequency Array (LOFAR): spatial
tomography of neutral gas at high redshift in 21cm emission. While the expected temperature 
fluctuations
due to unresolved radio point sources is highly uncertain, we point
out that free-free emission 
from the ionizing
halos that reionized the universe should define a minimal bound. This 
emission is likely to
swamp the expected brightness temperature fluctuations, making proposed
detections of the angular patchwork of 21cm emission across the sky unlikely to be viable. An alternative approach is to discern
the topology of reionization from spectral features due to 21cm emission along a pencil-beam
slice. This requires tight control of the frequency-dependence of the
beam in order to prevent foreground sources from contributing excessive
variance. We also investigate potential contamination by galactic and
extragalactic radio recombination lines (RRLs). These are unlikely to be
show-stoppers, although little is known about the distribution of RRLs
away from the Galactic plane. 
The mini-halo emission signal is always less than that of the IGM, making 
mini-halos unlikely to be detectable. If they are seen, it will be only
in the very earliest stages of structure formation at high
redshift, when the spin temperature of the IGM has not yet decoupled
from the CMB. 
\end{abstract}
\section{Introduction}

We have as yet no direct observational probes of neutral gas
beyond the epoch of reionization. One promising technique which has
attracted much attention is 21cm tomography of neutral hydrogen
\cite{scottrees,MMR,tozzietal}. 21cm emission and/or absorption from
neutral hydrogen at high redshift should exhibit angular
fluctuations as well as structure in redshift space. These
fluctuations are due to spatial
variations in the hydrogen density, ionization fraction, and spin
temperature, and may be detectable by future radio telescopes such
as the Square Kilometer Array (SKA)\footnote{see http://www.nfra.nl/skai} and the Low Frequency Array (LOFAR)\footnote{see http://www.astron.nl/lofar}. Although the
energy density in 21cm emission is about two orders of magnitude less
than the cosmic microwave background (CMB), the extreme smoothness of
the CMB both spatially and in frequency space would allow this signal to be
teased out. The expected brightness temperature fluctuations on
arminute scales is about two orders of magnitude larger than CMB
fluctuations. In principle, 21cm tomography would allow us to map out the
topology of reionization.  This could indirectly constrain the nature of the
ionizing sources, telling us whether these sources were faint and
numerous or bright and rare. It would thus be an invaluable tool for
probing the state of the intergalactic medium during an epoch when
traditional probes, such as Ly$\alpha$ absorption, fail (Gunn-Peterson absorption
fully saturates at hydrogen neutral fractions $x_{HI} \sim 10^{-4}$).

We present a study of possible foregrounds for these observations. On the arcminute
angular scales where the signal is expected to peak, detailed studies 
of the CMB (involving extrapolation from higher frequencies) tell us that
fluctuations in the galactic foreground emission are likely to be relatively
unimportant.  These galactic foregrounds are fairly well
understood and multi-frequency observations should enable us to
subtract out both the free-free and synchrotron components (see
\scite{shaver} for a detailed discussion). Unresolved extra-galactic radio
sources, on the other hand, could give rise to brightness temperature fluctuations which
would swamp the expected signal, as discussed by \scite{matteoetal}.
However, this model for
source counts was based on surveys with limiting flux densities $\sim
100$mJy at 150 MHz, which were then extrapolated down five orders of
magnitude past $\sim 1 \mu$Jy (the expected limiting point source
sensitivity of SKA). Therefore, as they acknowledge, this estimate of foreground
contamination is highly uncertain; a turnover in source counts at
lower flux levels could significantly reduce foreground contamination.  

We present a {\it minimal} estimate of the brightness temperature
fluctuations, based on free-free emission from the ionizing sources
that reionized the universe \cite{oh99}. We show that brightness
temperature fluctuations from these sources exceed the expected
signal, making it unlikely that the previously proposed angular 21cm 
tomography is feasible. We shall show that this estimate
is relatively model-independent and depends primarily on the
integrated ionizing emissivity.          

An inability to detect angular brightness temperature fluctuations need not render 21cm tomography studies powerless. Since the
foreground signal is expected to be smooth in frequency space, 21cm
spectral features along a pencil-beam slice, corresponding to
alternating patches of neutral and ionized hydrogen, could yield
invaluable information on the topology of reionization. However, this
relies heavily on the lack of foreground spectral contaminants. We
investigate two possible contaminants: spectral fluctuations in the foreground
due to the frequency-dependent beam, and radio recombination lines
from galactic and extragalactic ionized gas. The former can probably be dealt with if the frequency
dependence of the beam can be controlled and its side-lobes mapped
out. The latter is much more uncertain but is unlikely to be a show-stopper. If, in fact, it does turn out to be important,
it will yield the unexpected bonus of new information about the
clumping of ionized gas in our galaxy and in the universe as a whole.

In all numerical estimates, we
assume a $\Lambda$CDM cosmology where $(\Omega_{m},\Omega_{\Lambda},\Omega_{b}
h^{2},h,\sigma_{8 h^{-1}})=(0.3,0.7,0.019,0.7,0.9)$.

\section{Free-Free Emission from Ionizing Sources as a Foreground}

We begin by summarizing the properties of the expected signal; the
reader is referred to the literature \cite{scottrees,MMR,tozzietal}
for details. Consider a patch of
IGM, with spin temperature $T_{S}$, which fully fills the beam and has a
radial velocity width larger than the bandwidth of the radio
telescope. The differential brightness temperature between this patch
and the CMB is \cite{tozzietal}: 
\begin{equation}
\delta T_{b} \approx 9.0 {\rm mK} h^{-1} \left( \frac{ \Omega_{b}
  h^{2}}{0.02} \right) \redshiftnine^{1/2} \left( \frac{T_{S}-T_{CMB}}{T_{S}}
  \right).
\label{eqn:brightness_temperature}
\end{equation}
Of course, if ionized bubbles exist within this patch then the signal
will be diluted by the corresponding filling factor.

Initially,
$T_{S}=T_{CMB}$, but over time Ly$\alpha$ photons from the soft UV
background emitted by the first stars and quasars will couple the spin 
temperature of the gas to its kinetic temperature through the
Wouthuysen-Field effect. Since initially $T_{K} < T_{CMB}$, neutral hydrogen
will be visible for a brief period in absorption until the same
Ly$\alpha$ photons also heat the gas through recoil of the scattered
Ly$\alpha$ photons and make $T_{S} > T_{CMB}$. At this
point the signal is only visible in emission; from equation (\ref{eqn:brightness_temperature}) the brightness
temperature is roughly independent of the spin temperature. Hereafter, we
shall focus solely on the emission signal.

If linear theory is used to compute
expected density fluctuations, the brightness temperature fluctuations
peak on arcminute scales and have a rms value of $\langle \delta
T_{b}^{2} \rangle^{1/2} \approx 10 {\rm mK}$ for a $\Lambda$CDM
cosmology (see Fig 1 of \scite{tozzietal}). To give a sense of the
lengthscales involved, a bandwidth $\Delta \nu$ corresponds to a
comoving length $L \approx (1+z) c H(z)^{-1} \Delta \nu/\nu = 8.6
\redshiftnine^{-3/2} (\Delta \nu/1 {\rm MHz})  h^{-1} \, {\rm Mpc}$, where
$\nu=1.4/(1+z)$Ghz is the observation frequency, and an angular
diameter $\Delta \theta$ corresponds to a comoving transverse length $L=
\Delta \theta (1+z)/d_{A}(z) \approx 1.9 \left(\Delta \theta/1'
\right) h^{-1} \, {\rm Mpc} $ at z=9.      
 
The contribution of radio-loud AGN and radio galaxies at low flux
levels is highly uncertain. We therefore construct a minimal model of
the low-frequency radio background from free-free emission by ionizing
sources.  This background is {\it unavoidable} in the sense that a minimal
emissivity in ionizing photons is necessary to reionize the
universe. We present a brief summary here; refer to \scite{oh99} for
details. The free-free luminosity of an ionizing source can be
estimated from the fact that the free-free emissivity $\epsilon_{\nu}
\propto n_{e}^{2}$ is directly proportional to the
recombination rate and thus to the production rate of ionizing photons:
$L_{\nu}^{ff} \propto \langle n_{e}^{2} \rangle V \propto
\dot{N}^{recomb} \propto (1-f_{esc}) \dot{N}_{ion}$. We thus obtain:
\begin{equation}
L_{\nu}^{ff} \approx 1.1 \times 10^{27} \left( \frac{\dot{N}_{ion}}{10^{53}
{\rm photon \, s^{-1}}} \right) \left(\frac{1-f_{esc}}{0.9} \right)
{\rm erg \, s^{-1} Hz^{-1}}  
\end{equation}
where $\dot{N}_{ion}$ is the production rate of ionizing photons, and
$f_{esc}$ is the escape fraction of ionizing photons from the host
galaxy. For a Salpeter IMF with stars of solar metallicity,
$N_{ion}=10^{53} {\rm photons \, s^{-1}} \left( {\rm SFR}/1 {\rm M_{\odot}
yr^{-1}} \right)$. For a given star formation rate (SFR), stars with zero
metallicity can be one to two orders of magnitude more effective at producing ionizing
photons, since the effective temperature of these stars is higher
\cite{tumlinsonshull}, and the IMF of zero-metallicity stars
is often thought to be top-heavy. The escape fraction of ionizing photons is
observed to be $f_{esc} \sim 5 \%$ in the local universe \cite{leitherer}, and
thought to decline with redshift \cite{woodloeb,ricottishull},
although this is highly uncertain (for calculations arriving at the 
opposite conclusion, see \scite{fujitaetal}). The observed free-free flux is
then given by $S_{\rm ff}=L_{\nu}^{\rm ff}(1+z)/4 \pi d_{L}^{2}$, or:
\begin{equation}
S_{\rm ff} \approx 1.2 \redshiftnine^{-1} \left( \frac{\dot{N}_{ion}}{10^{53}
{\rm photons \, s^{-1}}} \right) \left(\frac{1-f_{esc}}{0.9} \right)
{\rm nJy}.
\label{eqn:flux}
 \end{equation}
Note that the uncertainty in the escape fraction of ionizing photons introduces uncertainties of at most
a factor of only a few in the free-free luminosity. Only in the unlikely 
scenario of virtually all
the ionizing photons escaping into the IGM without any
photo-electric absorption in the host ISM, would order of magnitude uncertainties creep in as $f_{esc} \rightarrow 1$. 

To compute the brightness temperature fluctuations due to these
sources, we need to construct a luminosity function. \scite{oh99}
followed the star formation model of \scite{haimanloeb}, who assumed
that some constant fraction of the gas $f_{*}$ in halos with virial
temperatures $T_{vir} > 10^{4}$K fragments to form stars, and that
each starburst lasts for $\sim 10^{7}$yr. The star formation
efficiency parameter $f_{*}$ is
tuned for consistency with the observed metallicity of the IGM at z=3, with
$f_{*} \sim 1.7-17\%$ corresponding to $Z \sim 10^{-3}-10^{-2}
Z_{\odot}$, respectively. With this prescription, $\dot{N}_{ion} \approx 2
\times 10^{53} \left( M_{halo}/10^{9} M_{\odot} \right) {\rm photons \,
s^{-1}}$. Press-Schechter theory can be used to compute the
abundance of halos and hence the source counts $dN/dS$. The power
spectrum of Poisson fluctuations can then be computed from:
\begin{equation}
C_{l}^{Poisson}= \int_{0}^{S_{c}} dS \frac{dN}{dS} S^{2}
\end{equation} 
where $S_{c}$ is the minimal flux above which point sources can be
identified and removed. The power spectrum due to clustering can be
computed from:
\begin{equation}
C_{l}^{clustering}= w_{l} I_{\nu}^{2}
\label{eqn:clustering}
\end{equation} 
where $w_{l}$ is the Legendre transform of the angular correlation
function of sources $w(\theta)$, and $I_{\nu}=\int_{0}^{S_{c}} dS
(dN/dS) S$ is the surface brightness of the radio background at
frequency $\nu$. The spatial correlation length can be computed from
linear theory assuming linear bias; the decrease in the linear growth
factor at high-redshift is increased by the number-weighted bias of
objects with $T_{vir} > 10^{4}$. This results in an angular correlation
function $w(\theta)=(\theta/\theta_{o})^{-0.8}$ with a nearly constant
angular correlation length $\theta_{o} \sim 2'$ independent of
redshift for $z > 3$ (see Fig. 5 of \scite{oh99}). The rms
temperature fluctuations can then be computed from
$T_{rms}=\left(l(l+1)C_{l}/ 4 \pi\right)^{1/2} \times c^{2}/2k_{B}
\nu^{2}$, where the frequency dependence of the temperature conversion
is appropriate in the Rayleigh-Jeans limit. This quadratic frequency dependence
implies that the population of free-free emitters only becomes
important at low frequencies. Thus, for instance, their existence does
not violate any present CMB distortion constraints. 

\begin{figure}
\psfig{file=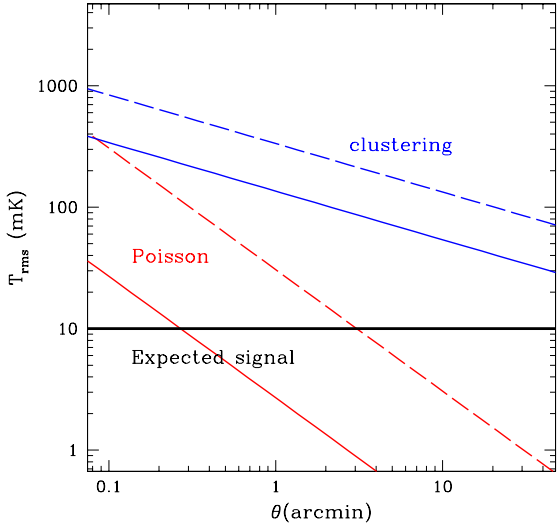,width=80mm}
\caption{Brightness temperature fluctuations induced by free-free
  emission from ionizing sources, compared with the expected 21cm
  signal, at a frequency of 150 MHz (corresponding to an observed
  redshift of $z=8.3$), and assuming a star formation efficiency
  corresponding to a Ly$\alpha$ forest metallicity of $Z=10^{-2.5}
  Z_{\odot}$ at $z=3$. Solid and dashed lines assume point source
  removal down to 20 nJy and $2 \mu$Jy respectively. The 21cm signal is likely
  to be swamped by fluctuations due to clustering of the free-free
  sources at all scales. The estimate for the clustering component is fairly robust
  and model independent (see text). Note that $T_{rms}^{cluster}
  \propto \Omega_{*}(z=3) \propto Z(z=3)$.}
\label{fig:free_free}
\end{figure}

What is the minimal flux $S_{c}$ above which point sources can be
identified and removed? The detector noise is:
\cite{rohlfswilson}:
\begin{equation}
S_{\rm inst}= \frac{2 k T_{\rm sys}}{A_{\rm eff} \sqrt{2 t \Delta
    \nu}} \sim 0.3 \left( \frac{\Delta \nu}{1 \, {\rm MHz}}
    \right)^{-1/2} \left( \frac{t}{10^{5} \, s} \right)^{-1/2} \,
    {\rm \mu Jy}
\end{equation}
where we have used $A_{\rm eff}/T_{\rm sys} = 2 \times 10^{8} {\rm
cm^{2} K^{-1}}$. On the other hand, the confusion noise is:
\begin{equation}
S_{\rm conf}=\left( C_{l}^{cluster}(S_{c}) \theta^{2}
\right)^{1/2} \approx 2 \left(\frac{S_{c}}{20 {\rm nJy}} \right)^{0.3}
\left( \frac{\theta}{0.3''} \right)^{1.6} {\rm nJy}
\end{equation}
Thus, for a 10$\sigma$ detection, the cutoff flux is $S_{c} \sim 20 \,
  {\rm nJy}$. Note that $S_{\rm conf} \propto \left( C_{l}^{cluster}(S_{c})
  \right)^{1/2} \propto S_{c}^{0.3}$ depends only
  weakly on the flux cutoff (see Fig. \ref{fig:free_free} and also
  Fig. 4 of \scite{oh99}). This is because $C_{l}^{cluster} \propto
  I_{\nu}^{2}(S_{c}) $, and from equation (\ref{eqn:flux}) the
  majority of sources which constitute
  the free-free background are generally fainter than $\sim 20 \,
  $nJy; removal of rare bright sources has little effect on the mean
  free-free background $I_{\nu}$. We have assumed unresolved point
  sources so that $\theta$ is given by the beam resolution; the
  confusion noise becomes correspondingly larger if the source is extended. 

Confusion noise, rather than instrumental noise, is the
  limiting factor in source identification and removal for free-free
  sources. The free-free flux is largely frequency-independent. If
  instrumental noise were the limiting factor, point sources could be
  identified and removed with higher frequency observations, for which
  instrumental noise is much lower (for instance, in 10  days a
  5$\sigma$ detection of a 16 nJy source at 2 GHz is possible, for a
  bandwidth $\Delta \nu=1$ GHz). However, since
  confusion noise is frequency independent, once the confusion
  limit is reached no further source removal is possible, even with
  multi-frequency observations. 
 
We compute the Poisson and clustering contributions to the free-free
background with the above model. The results are
displayed in Figure \ref{fig:free_free}. The 21cm signal is likely
  to be swamped by Poisson fluctuations at small scales and by
  fluctuations due to clustering of free-free sources at all
  scales. This calculation should be
compared with that of \scite{matteoetal}, who use a power law extrapolation of
observed number counts to compute the same quantities. Although we
arrive at similar conclusions, below we argue that these estimates (in
particular the temperature fluctuations due to the clustering
component) are significantly more robust.

The results obviously depend on the specifics of the star formation
model assumed. Since this is highly uncertain, we highlight
here the most robust features of the calculation, which are largely
model independent. The number counts in the Poisson signal are dominated by rare
bright objects just below the detection threshold $S_{c}$, and can
vary widely depending on the shape of the luminosity function at these
low flux levels. We therefore disregard this term as it is excessively
model dependent. On the other hand, the clustering term, equation (\ref{eqn:clustering}) is
significantly more robust. Our estimate of $\theta_{o} \sim 2'$ is
likely a minimal estimate of the angular clustering strength since it
is dominated by the bias of halos at the threshold
virial temperature $T_{vir} \sim 10^{4}$K, which are most abundant.
Star formation in lower virial temperature halos, which cannot cool by atomic line cooling, is expected to be extremely inefficient; see
e.g. \scite{barkanaloeb} for a general review. If, in fact, star
formation is more efficient in deeper potential wells (e.g., since feedback processes are less devasting there), the
luminosity weighted correlation function will be more heavily weighted
towards rare massive halos, which are more highly biased. Thus, the
$w_{l}$ term is generally a minimal estimate. The surface
brightness term $I_{\nu}^{2}$ is independent of the model for the luminosity
function and depends only on the overall normalization of the comoving luminosity density. We can see this from the equation of cosmological radiative transfer \cite{Peebles93}:
\begin{equation}
I(\nu_{o})= \frac{c}{4 \pi} \int_{0}^{\infty} dz \frac{dt}{dz} \frac{\epsilon[\nu_{o}(1+z)]}{(1+z)^{3}} 
\end{equation} 
where $\epsilon(\nu)$ is the comoving emissivity. Since $\epsilon \propto \dot{\Omega}_{*} (1-f_{esc})$ (from
equation \ref{eqn:flux}), the radio surface
brightness $I(\nu_{o}) \propto \Omega_{*}(1-f_{esc})/(1+\bar{z})^{3}$, where
$\bar{z}$ is the median redshift at which $\sim 50 \%$ of all stars
have formed. Thus, a given comoving stellar density $\Omega_{*}$ directly
and robustly implies a minimal free-free surface brightness on the sky (the
dependence on the star formation history $\dot{\Omega}_{*}(z)$ is weak,
since most stars formed at late times). In our calculation we have conservatively included only sources at $z > 3$ (which are too faint to be identified and removed) and simply normalized to
$\Omega_{*} (z=3) = 6 \times 10^{-3} \left( \bar{Z}/10^{-2.5} Z_{\odot}
\right) \Omega_{b}$, where $\bar{Z}$ is the mean metallicity of the IGM at $z
\sim 3$, and we assume a Salpeter IMF where $\sim 1 M_{\odot}$ of
metals form for every $\sim 100 M_{\odot}$ of stars formed. As previously noted, because the  
majority of sources are significantly fainter than the threshold flux for
point source removal $S_{c}$, the mean surface brightness depends only weakly
on $S_{c}$ (see Fig. 4 of \scite{oh99}). Since both the estimates for
the clustering strength $w_{l}$ and the mean surface brightness
$I_{\nu}$ are minimal and robust, we conclude that our estimate for
$C_{l}^{cluster}$ constitutes a robust {\it minimal} estimate for
angular temperature fluctuations on the sky. These fluctuations will swamp the
expected brightness temperature fluctuations due to 21cm emission by
at least an order of magnitude.

\section{Foregrounds for Spectral Measurements}

Even if brightness temperature fluctuations due to 21cm emission 
are not detectable,
spectral variation along a pencil-beam might
still be detected. This relies on the fact that foreground sources
should in general have smooth power-law continuum spectra, and
averaging over many sources with different spectral indices and
spectral structure will yield a smooth power-low foreground (see
\scite{shaver} for simulations of this) which would still allow
detection of spectral structure
corresponding to alternating regions of neutral and ionized hydrogen. 
To give an idea of scale, an ionized 'bubble' created by a
source forming stars at a rate $\sim 2 {\rm M_{\odot} yr^{-1}}$ for
$\sim 10^{7}$yr has a diameter
$\sim 1.4 [(1+z)/10]^{-1} h^{-1} {\rm Mpc}$ comoving, while a
bandwidth $\Delta \nu$ spans a comoving length $L \approx 8.6
\redshiftnine^{-3/2} (\Delta \nu/1 {\rm MHz})  h^{-1} \, {\rm
  Mpc}$. Of course, as reionization proceeds the ionized regions
expand in size and neutral regions contract. As \scite{matteoetal} point out,
observing with 2 MHz resolution at $\nu_{o}=150$MHz with an error of $\Delta \beta \sim
0.05$ in the measured foreground spectrum spectral slope should allow one to
distinguish a signal that is $[(\nu/\nu_{o})^{\Delta \beta} -1] \sim 6
\times 10^{-4}$ times smaller than the foreground. Here, we discuss
two potential foregrounds for such spectral measurements: beam smearing
of foreground sources and radio recombination lines. 

The radio telescope does not sample exactly the same patch of sky at
all frequencies, since the beam-size and sidelobes are frequency
dependent. This would by itself naturally introduce frequency space
fluctuations in the measured brightness temperature as additional
sources fill the beam, an effect of order:   
\begin{eqnarray}
\left( \frac{\Delta T}{T} \right)_{fg} &\sim& \frac{\Delta
  \Omega}{\Omega} \sim 2 \frac{\Delta \theta}{\theta} \sim 2 \frac{\Delta
  \nu}{\nu} \\ \nonumber
&\sim& 1.4 \times 10^{-2} \left( \frac{\Delta \nu}{2 {\rm
  MHz}} \right) \redshiftnine 
\end{eqnarray}
Since we have seen that the clustering foreground $\langle T^{2}_{fg}
\rangle^{1/2} \geqsim 10 \langle T^{2}_{21 cm} \rangle^{1/2}$, the
noise in frequency space introduced by beam-size variation will be
$\delta T_{fg} \sim {\rm few} \times 0.1 \langle T^{2}_{21 cm} \rangle^{1/2}$,
comparable to temperature fluctuations due to 21cm emission. An exception
may be in the very early/late stages of reionization, when the size
of ionized/neutral patches is small; the beamsize does not vary
signficantly over the small bandwidth $\Delta \nu$ necessary to detect spectral
feaures. Otherwise, it would be necessary to develop scaled beams and controlled sidelobes, which are capable of sampling exactly the same
patch of sky at all frequencies. Note that frequency calibration of
the telescope for pencil beam tomography is considerably
more challenging than frequency calibration for detecting an all-sky
signal, such as that produced at the tail end of reionization
\cite{gnedin}.  In the case of an all-sky signal, one could simply 
average the results over many
independent patches of sky, as well as perform a differencing
experiment off the moon, which blocks emission behind it \cite{shaver}. 

Another possible source of structure in frequency space is the signal 
from radio
recombination lines (RRLs). These lines have been observed both in emission
and absorption at a wide range of frequencies, although generally only against bright sources in
the Galactic plane (see \scite{gordon} for a comprehensive review). The frequencies of the hydrogen lines are:
\begin{equation}
\nu \approx 153 \Delta n \left( \frac{n}{350} \right)^{-3} {\rm MHz}. 
\end{equation}
The important question is whether RRLs will provide a significant
source of contamination on lines of sight outside the Galactic
plane. There have been few RRL line searches away from bright sources
outside the Galactic plane, and in any case surveys for radio
recombination lines generally have low-frequency detection limits 
above the strength of the 21cm features we seek. We can
place a useful limit using the fact that the observed brightness 
of both RRL and H$\alpha$ emission depends on the emission measure
$EM=\int ds n_{e}^{2}$, and that optical surveys for H$\alpha$ have
much lower limiting sensitivities. Fabry-Perot surveys have detected
H$\alpha$ emission from every Galactic latitude, with minimal values
$0.25-0.8$ Rayleighs toward the Galactic pole \cite{reynolds}. The
emission measure corresponding to an observed H$\alpha$ intensity
$I_{\alpha}$ (in Rayleighs) is: $EM(H \alpha)=2.75 T_{4}^{0.9}
I_{\alpha} {\rm cm^{-6} pc}$. Based
on the width of the H$\alpha$ line, \scite{reynolds} estimates $T \sim
8000$K. 
Given the density, temperature and size of a region, as well as an
estimate
of the linewidth and a frequency to observe at, one can also calculate the
recombination line optical
depth \cite{shaver75}:
\begin{displaymath}
\tau_{L} =
\frac{518 EM}{\nu \Delta V_{L} T_{e}^{5/2}}
b_{n} \left[ 1-20.8
  \frac{T_{e}}{\nu}
  \frac{d \ln b_{n}}{dn}
\right],
\end{displaymath}
where $EM$ is the emission measure in cm$^{\textrm{-}6}$pc defined by $n_{e}^{2\
}l$,
$\nu$ is the line center frequency, $V_{L}$ is the line width in km/s,
$T_{e}$
is the electron temperature, and $b_{n}$ is the departure coefficient,
which takes into account departures from local thermodynamic
equilibrium (LTE). For LTE, $b_{n} \equiv 1$. For an unresolved line,
the velocity width is simply: $\Delta V_{L} = c (\Delta \nu/\nu)$,
where $\Delta \nu$ is the observational bandwidth. Assuming $I_{\alpha}=0.25 R$ (which corresponds to $EM=0.5 {\rm
  cm^{-6} pc}$) and $T_{L}=\tau_{L} T_{e}$, we can employ the optical depth
calculation to estimate the hydrogen RRL temperature:
\begin{equation} 
T_{L} = 1.3 \times 10^{-6} \left( \frac{T_{e}}{8000 K} \right)^{-3/2}
\left( \frac{EM}{0.5 {\rm cm^{-6} pc}} \right) \left( \frac{\Delta
  \nu}{1 {\rm MHz}} \right)^{-1} \, {\rm K}  
\end{equation}
which is negligible. Note, however, that this estimate assumes
thermodynamic equilibrium. Particularly for lower frequency lines,
stimulated emission may become important.

Carbon radio recombination lines are another possible source of
contamination. They have been detected in the frequency range 34-325
MHz toward Cass A in the Galactic plane \cite{Payne}, and also in
a 327 MHz survey centered at $b=14^{\circ}$ \cite{roshi}. Optical depths are
generally $\tau \sim 10^{-3}$ and they are thought to arise in dense,
cold ($T_{e} \sim 20-200$K), partially ionized regions where non-LTE
effects are important. Carbon RRL emission is generally confined to $b
< 3^{\circ}$ and their intensity out of the plane is probably small
\cite{roshi}.   

Detections of radio recombination lines from extragalactic sources
have generally been confined to bright starburst galaxies and the
intensity appears to be correlated with the star formation rate 
\cite{phookun}. They are unlikely to be significant
contaminants. One can show that for reasonable assumptions about the
excitation parameter only nearby sources ($D \le 10$Mpc)
will be detectable in RRL emission if spontaneous emission is
responsible \cite{shaver78}. It was thought that
stimulated emission could allow observation of much more distant
sources \cite{shaver78}, but that has not turned out to be the case. 

Extragalactic RRLs could be a foreground in
the search for 21cm {\it absorption} against radio-loud sources at
high redshift \cite{carilli,furlanetto}. Since high-redshift objects
are in general much denser $n_{e}^{2} \propto (1+z)^{6}$, their
emission measures are much higher. Consider an isothermal disk at
temperature $T_{gas}$ embedded in a halo with virial temperature $T_{vir}$.
One can solve for the density profile of such a disk, $n \propto
{\rm exp}(-r/R_{d}) {\rm sech^{2}}(z/z_{o})$ (\scite{woodloeb}; see also
\scite{oh_haiman}), and the emission measure of the disk will be $EM(r)
\approx \int dz n^{2}(r,z)$. For a disk seen face-on, the emission measure 
becomes a function of the radial distance $r$ from the center:
\begin{eqnarray}
\nonumber
&& EM(r) \approx 2 \times 10^{9} 
\left( \frac{f_{\rm d}}{0.5} \right)^{3} 
\left( \frac{T_{\rm gas}}{10^{4} \, {\rm K}} \right)^{-1} 
\left( \frac{T_{\rm vir}}{5 \times 10^{4} \, {\rm K}} \right)^{3/2} \\
&&
\times
\left( \frac{\lambda}{0.05} \right)^{-6} 
\left( \frac{1+z}{10} \right)^{9/2} {\rm exp}\left(-\frac{2 r}{R_{\rm d}} \right) \ {\rm cm^{-6} pc}
\label{eqn:emission_measure}
\end{eqnarray}
where $f_{d}=(M_{disk}/M_{halo})/(\Omega_{b}/\Omega_{o})$ is the fraction of baryons in the disk, $\lambda$ is the
spin parameter and $R_{d} \approx \lambda r_{vir}/\sqrt{2}$. The
optical depth at line center for a $\Delta n=1$ RRL will be:
\begin{equation}
\tau_{L}= 2 \times 10^{-2} \left( \frac{T_{e}}{10^{4} K}
\right)^{-5/2} \left( \frac{EM}{10^{8} {\rm cm^{-6} pc}} \right)
\left( \frac{\Delta \nu}{1 {\rm MHz}} \right)^{-1},
\end{equation}  
assuming local thermodynamic equilibrium and that the line is unresolved. This could
be considerably enhanced if stimulated emission and/or non-LTE effects become important (which is
likely since one is looking along the line of sight to a radio-bright
source). This optical depth is comparable to the central optical depths due to 21cm absorption
by the IGM \cite{carilli} or mini-halos/mini-disks \cite{furlanetto}, which
are of order $\tau \sim 10^{-2}$. If one uses Press-Schechter theory
to calculate abundances of halos with $T_{vir} > 10^{4}$K, a random line of sight
would intersect $\sim {\rm few} \times 0.1$ disks per redshift
interval, for the redshifts of interest \cite{furlanetto}. Even a
single disk along the line of sight would introduce a plethora of RRL features in emission and/or
absorption. This would make the task of picking out the spectral
features due to 21cm absorption much more difficult. On the other hand, such observations would also be an invaluable probe of gas clumping
at high redshift. However, this is a rather speculative scenario, and we shall not
comment on it further. 

In general, radio recombination lines are unlikely to be
significant contaminants. If they are, they will
have to be removed with observations of higher spectral resolution
(using the fact that they occur at known frequency intervals for
identication and removal). In many ways their detection could be an 
unexpected
bonus, as RRLs have the potential to teach us a great deal about the
clumping of ionized gas both within our Galaxy and in the universe as a whole. 

\section{Detection of Mini-Halos}

Recently, it has been proposed that mini-halos with $T < 10^{4}$K,
which are not collisionally ionized but which have gas densities 
sufficiently high to collisionally couple the spin temperature to the 
kinetic temperature, may be detectable in
emission \cite{mh1,mh2}. For $T_{S} \gg T_{CMB}$, the flux is
independent of the spin temperature and depends only on the HI
mass. Although the signal from an individual mini-halo is very faint,
the combined signal from many mini-halos within a sufficiently large 
comoving volume may be detectable.
We point out that because
$S_{\nu} \propto M_{HI}$, the flux from mini-halos will always be
swamped by the flux from the neutral IGM, as the HI mass in the IGM
is larger. The ratio of fluxes is given by the relative fraction of
HI in collapsed halos with $T_{vir} < 10^{4}$K:
\begin{equation}
\frac{S_{\nu}^{halos}}{S_{\nu}^{IGM}} \approx
\frac{\Omega_{HI,halos}}{\Omega_{HI}} \approx
\frac{1}{\rho_{m}}\int_{M_{low}}^{M_{high}} dM \frac{dN}{dM}{M}
\label{eqn:ratio_fluxes}
\end{equation}
where $dN/dM$ is the halo abundance given by Press-Schechter theory, $M_{high}=10^{8} [(1+z)/10]^{-3/2} {\rm M_{\odot}}$ is the mass of
halos with $T_{vir}=10^{4}$K, and $M_{low}$ is the baryonic Jeans mass
below which gas cannot accrete onto halos. If there is no heating of
the IGM, $ M_{low}$ is given by the cosmological Jeans mass $M_{J}=8 \times
10^{3} [(1+z)/10]^{3/2} {\rm M_{\odot}}$, where the temperature of the
IGM $T_{IGM}=1.7 [(1+z)/10]^{2}$K is simply set by the amount of 
adiabatic cooling
since decoupling from the CMB. In reality it is likely that the gas will
be heated up to higher temperatures by soft X-ray
and Ly$\alpha$ photon recoil heating
from the first sources of light, as well as by free-free
emission from collapsing structures \cite{MMR}. This will suppress 
the accretion of gas onto the smallest mini-halos.   

\begin{figure}
\psfig{file=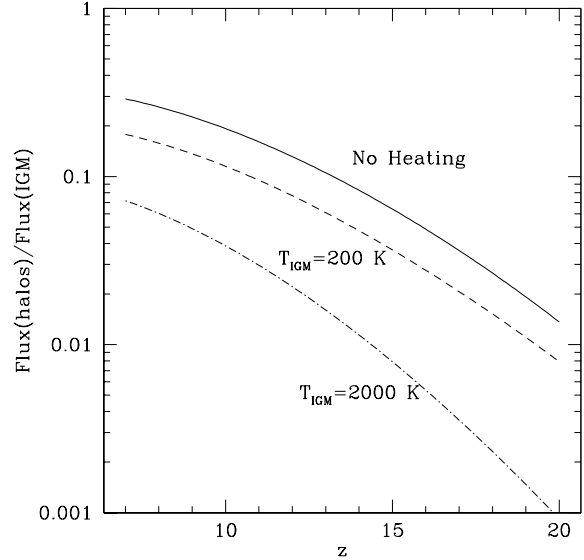,width=80mm}
\caption{Relative 21cm flux from mini-halos compared to IGM in a
  neutral patch which fills the beam, assuming $T_{S} \gg T_{CMB}$ for
  both. Since $S_{\nu}^{halos}/S_{\nu}^{IGM} = \Omega_{halos}/\Omega_{b}$,
  the 21cm emission from the IGM will always be greater except in the
  earliest stages of structure formation when the spin temperature of
  the IGM is still coupled to that of the CMB. The relative mass fraction (and hence
  21cm flux) in mini-halos decreases still further if the IGM is
  heated to 'warm' temperatures (e.g. $T_{IGM} \sim {\rm 200K, 2000 K}$ as
  shown here) by soft X-rays or atomic recoil from
  Ly$\alpha$ photon scattering, which increases the baryonic Jeans
  mass and suppresses accretion onto mini-halos.}
\label{fig:minihalo_mass_fraction}
\end{figure}

We plot the ratio of fluxes from equation (\ref{eqn:ratio_fluxes}) in
 Fig (\ref{fig:minihalo_mass_fraction}). Even if heating is negligible, this
 ratio is at most a ${\rm few} \times 0.1$. This may be
 amplified in high density peaks by a factor of no more than a few due to clustering bias
 \cite{mh2}. Thus, the flux from mini-halos only dominate in two
 limits: (i) in reionized regions, for self-shielded mini-halos which have not yet been photo-evaporated, and (ii) very early in
 the structure formation process, when the background radiation field
 is low and the spin temperature of the IGM is still coupled to that of
 the CMB. The former case is unlikely to be detectable as the
 mini-halos contribute a small signal superimposed on the much larger
 fluctuating signal of alternating neutral and ionized
 regions of the IGM. The latter may be detectable provided the
 mini-halos indeed remain neutral and do not form stars via ${\rm H_{2}}$
 cooling: even a single star in a halo is capable of photo-ionizing and
 photo-evaporating much of the gas. 

It is worth checking if this is possible. From \scite{HaimanAbel} (see
 their Fig. 6), the level
 of background UV flux capable of photo-dissociating ${\rm H_{2}}$ in
 mini-halos and preventing star formation is:
\begin{equation}
J_{diss} \sim 10^{-3}-10^{-2}
 J_{21} \nonumber
\end{equation}
 in the redshift range z=10-20 where $J_{21}$ (in units of $10^{-21}
 {\rm erg s^{-1} cm^{-2} Hz^{-1} sr^{-1}}$) is the UV flux in the Lyman-Werner bands
 $11.2-13.6$eV. On the other hand, the critical thermalization flux of
 Ly$\alpha$ photons (which consists of redshifted photons from the
 Lyman-Werner bands) required to decouple the spin temperature of the
 IGM from the CMB and drive $T_{S} \rightarrow T_{K}$ is \cite{MMR}:
\begin{equation} 
J_{therm} \approx 0.9 J_{21} \redshiftnine. \nonumber
\end{equation}
Since $J_{diss} \ll
 J_{therm}$, there certainly exists a period in the history of the IGM
 where gas in mini-halos cannot cool to form stars, but the IGM does not emit in
 21cm radiation. In this case, the halos would give rise to 
 a period in which 21cm is detectable in emission before it is seen in
 absorption when $T_{S,IGM} < T_{CMB}$. 21 cm radiation would once again
be detectable in
 emission due to heating when $T_{S,IGM} \gg T_{CMB}$. The duration of each of
 these epochs can be used to time the rise of the radiation field. Of
 course, this is all assuming that the noise from foregrounds
 previously discussed can be overcome for the smaller signal levels expected from mini-halos. Note that a period of early reionization would wipe out the
 mini-halo population by raising the baryonic Jeans mass \cite{cen}.
 Also, early star formation would provide trace metal contamination,
 and if $Z \geqsim 10^{-3} Z_{\odot}$ \cite{Bromm}, the gas in mini-halos would
 be
 able to cool via metal line cooling and form stars. If the mini-halo signal is detectable in
 spectral pencil-beam measurements, it is preferable to have as
 small a bandwidth as possible ($\sim 0.1$MHz) to maximize intensity
 fluctations due to the varying number of halos within a bandwidth
 along the line of sight.  

\section{Conclusions}

We have considered a number of foregrounds for 21cm
emission studies. We perform a robust minimal estimate of 
temperature fluctuations across the sky due to free-free emission by
ionizing sources assuming only: (i) a number-weighted clustering
bias with a cutoff for halos with $T_{vir}< 10^{4}$K, (ii) a
star formation history normalized to the metallicity of the z=3
Ly$\alpha$ forest, and (iii) an escape fraction of ionizing photons into
the IGM of no more than $\sim {\rm few} \times 0.1$. We conclude
that the clustering component of temperature fluctuations will swamp
the expected angular brightness temperature fluctations due to 21cm emission
during the cosmological Dark Ages by at least an order of
magnitude. Performing 21cm tomography by detecting spectral 
features along a
single line of sight requires developing scaled beams which sample
exactly the same patch of sky at all frequencies. Otherwise,
excess variance in foreground sampling due to the frequency dependence
of the beam will swamp the signal. We also discuss radio recombination
lines as possible contaminants. It is unlikely that this will be a
problem for surveys outside of the Galactic plane, although this is 
still fairly uncertain.   

Detection of a 21cm spectral signature in absorption against a 
high-redshift radio loud
source \cite{carilli,furlanetto} will not suffer from the foregrounds
discussed here, except for the rather speculative possibility that
intervening high-redshift disks could give rise to a plethora of 
obscuring radio recombination lines. Of
course, this 'foreground' would itself be a marvellous window on the
high-redshift universe! The main difficulty of absorption studies, of
course, is whether such bright radio sources exist at early times,
when structure formation is in its infancy.    

\section*{acknowledgments}
KM thanks Mark Gordon, Bill Erickson, Peter Shaver, Harry Payne, Miller Goss,
and Rod Davies for helpful correspondence and 
the Caltech Summer Undergraduate Research Fellowship (SURF)
office at Caltech for financial support. SPO was supported by NSF 
grant AST-0096023.


\begin{thebibliography}{}

\bibitem[Barkana \& Loeb <2001>]{barkanaloeb} Barkana, R., \& Loeb, A., 2001,
  Phys. Rep., 349, 125

\bibitem[Bromm et al <2001>]{Bromm} Bromm, V., Ferrara, A., Coppi,
  P.S., \& Larson, R.B. 2001, MNRAS, 328, 969 

\bibitem[Carilli, Gnedin \& Owen <2002>]{carilli} Carilli, C.L.,
  Gnedin, N.Y., \& Owen, F., 2002, ApJ, 577, 22

\bibitem[Cen <2002>]{cen} Cen, R., 2002, ApJ, submitted, astro-ph/0210473

\bibitem[Di Matteo et al <2002>]{matteoetal} Di Matteo, T., Perna, R.,
  Abel, T., \& Rees, M.J., 2002, ApJ, 564, 576

\bibitem[Fujita et al <2002>]{fujitaetal} Fujita, A., Martin, C.L.,
Mac Low, M.-M., \& Abel, T., ApJ, submitted, astro-ph/0208278

\bibitem[Furlanetto \& Loeb <2002>]{furlanetto} Furlanetto, S.R., \&
  Loeb, A., 2002, ApJ, 571, 1

\bibitem[Gordon \& Sorochenko <2002>]{gordon} Gordon, M.A., \&
  Sorochenko, R.L., 2002, {\it Radio Recombination Lines: Their
  Physics and Astronomical Applications}, Kluwer, Dordrecht

\bibitem[Gnedin \& Ostriker <1997>]{gnedin} Gnedin, N.Y., \& Ostriker,
  J.P., 1997, ApJ, 486, 581

\bibitem[Haiman \& Loeb <1997>]{haimanloeb} Haiman, Z., \& Loeb, A.,
  1997, ApJ, 483, 21 

\bibitem[Haiman, Abel \& Rees <2000>]{HaimanAbel} Haiman, Z., Abel, T., \& Rees,
  M.J., 2000, ApJ, 534, 11

\bibitem[Iliev et al <2002a>]{mh1} Iliev, I.T., Shapiro, P.R.,
  Ferrara, A., \& Martel, H., 2002a, ApJ, 572, L123

\bibitem[Iliev et al <2002b>]{mh2} Iliev, I.T., Scannapieco, E.,
  Martel, H., \& Shapiro, P.R., 2002b, MNRAS, submitted, astro-ph/0209216

\bibitem[Leitherer et al <1995>]{leitherer} Leitherer, C., et al 1995, ApJ, 454, L19

\bibitem[Madau, Meiskin \& Rees <1997>]{MMR} Madau, P., Meiskin A., \&
Rees, M.J., 1997, ApJ, 475, 429  

\bibitem[Oh <1999>]{oh99} Oh, S.P., 1999, ApJ, 527, 16

\bibitem[Oh \& Haiman <2002>]{oh_haiman} Oh, S.P, \& Haiman, Z., 2002, ApJ, 569, 558

\bibitem[Payne et al <1989>]{Payne} Payne, H.E., Anantharamaiah, K.R.,
  \& Erickson, W.C., 1989, ApJ, 341, 690 

\bibitem[Peebles <1993>]{Peebles93} Peebles, P.J.E., 1993, Principles
  of Physical Cosmology, Princeton University Press, Princeton

\bibitem[Phookun et al <1998>]{phookun} Phookun, B., Anantharamaiah,
  K.R., \& Goss W.M., 1998, MNRAS, 295, 156

\bibitem[Reynolds <1990>]{reynolds} Reynolds, R.J., 1990, in S. Bowyet
  \& C. Leinert (eds), {\it Galactic and Extragalactic Background
  Radiation, Proc. IAU Symp. No. 139}, Kluwer, Dordrecht, p. 157

\bibitem[Ricotti \& Shull <2000>]{ricottishull} Ricotti, M., \&
  Shull, J.M., 2000, ApJ, 542, 548
 
\bibitem[Rohlfs \& Wilson <1996>]{rohlfswilson} Rohlfs, K., \& Wilson,
  T.L., 1996, Tools of Radio Astronomy (Berlin:Springer)

\bibitem[Roshi et al <2002>]{roshi} Roshi, D.A., Kantharia, N.G., \&
  Anantharamaiah, K.R., 2002, A\&A, 391, 1079

\bibitem[Scott \& Rees <1990>]{scottrees} Scott, D., \& Rees, M.J.,
1990, MNRAS, 247, 510

\bibitem[Shaver <1975>]{shaver75} Shaver, P.A., 1975, A\&A, 43, 465

\bibitem[Shaver <1978>]{shaver78} Shaver, P.A., 1978, A\&A, 68, 97

\bibitem[Shaver et al <1999>]{shaver} Shaver, P.A., Windhorst, R.A.,
Madau, P., \& de Bruyn, A.G., 1999, A\&A, 345, 380

\bibitem[Tozzi et al <2000>]{tozzietal} Tozzi, P., Madau, P., Meiskin,
A., \& Rees, M.J., 2000, ApJ, 528, 597

\bibitem[Tumlinson \& Shull <2000>]{tumlinsonshull} Tumlinson, J., \&
  Shull, M.J. 2000, ApJ, 528, L65

\bibitem[Wood \& Loeb <2000>]{woodloeb} Wood, K., \& Loeb, A., 2000, ApJ, 545, 86
 
\end{thebibliography}
\end{document}